\documentclass[UTF8,a4paper]{article}
\usepackage{amsmath}
\usepackage{graphicx}
\usepackage{subfigure}
\usepackage{epstopdf}
\usepackage{geometry}
\usepackage{ulem}
\usepackage{indentfirst}
\usepackage{amsfonts,amssymb,dsfont}
\usepackage{setspace}
\usepackage{threeparttable}
\usepackage{amsmath,mathtools,amsthm}
\usepackage{array}
\usepackage{extarrows}
\usepackage{upgreek}

\makeatletter
\renewcommand*\env@matrix[1][\arraystretch]{%
  \edef\arraystretch{#1}%
  \hskip -\arraycolsep
  \let\@ifnextchar\new@ifnextchar
  \array{*\c@MaxMatrixCols c}}
\makeatother
\begin{document}


\title{\bf Anomalous Rabi oscillation and related dynamical polarizations under the off-resonance circularly polarized light}
\author{Chen-Huan Wu
\thanks{chenhuanwu1@gmail.com}
\\Key Laboratory of Atomic $\&$ Molecular Physics and Functional Materials of Gansu Province,
\\College of Physics and Electronic Engineering, Northwest Normal University, Lanzhou 730070, China}

\maketitle
\vspace{-30pt}
\begin{abstract}
\begin{large}

We investigate the photoinduced effect to the silicene, which is a topological insulator, by the circularly polarized light in off-resonance regime
with a frequency much larger than the critical value (also
much larger than the frequency about the particle-hole pair creation),
and with a perpendicular electric field.
The anomalous Rabi frequency which is a non-linear optical, arised by the off-resonance circularly polarized light.
The temporal behavior of the pseudospin, valley, and spin degrees of freedom, which are momentum- and quasienergy-dependent,
are explored.
The anomalous Rabi oscillation is also related to the photoinduced topological phase transition between the topological trivial state with zero Chern number
and gapped edge state and the topological nontrivial state with nonzero Chern number and gapless edge state.
The off-resonance laser can also induce the topological phase transition by manipulating the energy band structure,
rather than excite the atoms to the high quantum-number states like the resonance light.
The exchange between the radiation driving field and the two-component dynamical polarizations with the dipole oscillation,
plays a important role in the determination of the out-of-plane spin polarization and the motion of the center of mass,
which can induced a collapse-and-revival pattern under a certain condition.
The Rabi oscillation observed in the motion of the center of mass in a laser-induced harmonic potential
can be used to detect the time evolution of the atom polulation.
Our results can also be applied to the other two-dimension low-energy Dirac models or the surface of the three-dimension topological insulators,
and even the weyl semimetal with the photoinduced topological phase transition.

\end{large}

\end{abstract}
\begin{large}
\section{Introduction}

Silicene, 
the silicon version of the graphene,
has attachted much attentions both experimentally
and theoretically since it's successfully synthesized
together with it's bilayer form or nanoribbon form\cite{Feng B},
and it has the properties of both the topological insulator (TI) and semimetal,
which provides possibility for the abundant phase transitions\cite{Ezawa M}.
The low-energy dynamics of silicene can be well described by the Dirac-theory.
The silicene is also a $3p$-orbital-based materials with the noncoplanar low-buckled (with a buckled diatance 
$\overline{\Delta}=0.46$ \AA\ due to the hybridization between the 
$sp^{2}$-binding 
and the $sp^{3}$-binding (which the bond angle is $109.47^\text{o}$) and 
that can be verified by thr Raman spectrum 
which with the intense peak at 578 cm$^{-1}$ larger than the planar one and the $sp^{3}$-binding one
\cite{Tao L},
and thus approximately forms two surface-effect (like the thin ferromagnet matter) lattice structure.
The bulked structure not only breaks the lattice inversion symmetry,
but also induce a exchange splitting between the upper atoms plane and the lower atom plane and thus forms a emission geometry which allows the
optical interband transitions, which for the graphene can happen only upon a ferromagnetic (FM) substrate.
The FM or antiferromagnetic (AFM) order can be formed in monolayer silicene by the magnetic proximity effect
that applying both the perpendicular electric field and in-plane FM or AFM field.

The degrees of freedom about the spin, valley, and sublattice of silicene are detected in this article.
Unlike the dynamical polarization due to the scattering by the charged impurities as we have explored\cite{Wu C H2,Wu C H3},
the out-of-plane dynamical polarization of spin, valley, and sublattice detected here exhibit high-frequency oscillation which 
related to the optical frequency of the applied light.
The manipulation of these degrees of freedom are important for the developing of spintronics and valleytronics
and even the pseudospintronics.
It's also been found that, the magnetic impurity scattering in a unbiased bilayer system\cite{Min H}
will leads to the broken of magnetic order by the in-plane pseudospin component,
while the out-of-plane component (mainly due to the charge transfer between the two layers)
is much larger than the in-plane one due to the strong exchange.

\section{Dynamic in off-resonance regime}
It's been found that the photoinduced topological phase transition\cite{Ezawa M} 
and the pseudospin redistribution\cite{López A} is possible by appliying the circularly polarized light
which can effectively manipulates the charge carriers and modulate the band gap.
In this paper we mainly focus on the circularly polarized light which can be realized by,
e.g., applying two time-dependent linearly polarized laser field ${\bf E}(t)=E(t)\hat{z}$ 
which are in different polarization directions that perpendicular to each other.
For silicene, we have found that the critical frequency for the optical resonance as well as the first-order-process is 1200 THz
($\approx 3t$ where $t=1.6$ eV is the nearest hopping parameter in the monolayer silicene).
Below this frequency, the first-order-process like the optical absorbtion $A_{op}(\omega)=(1/\epsilon_{0}c){\rm Re}[\sigma(\Omega)]$,
where $\epsilon_{0}$ is the permittivity of the vacuum and $\sigma$ is the optical frequency $\Omega$-dependent optical conductivity,
would happen,
while at the off-resonance regime whose frequency is much higher that this critical frequency, 
it's dominated by the second-order-process, like the photon-coupling and the optical reflection,
and the optical absorbtion is absence and instead, the virtual photon process emerges.
In off-resonance region, 
the Rabi oscillation as a nonlinear optics which is closely related to the 
pseudospin polarization can be studied by using the asymptotic rotating wave approximation (ARWA; or the Floquet approximation),
which is a frequency much larger than the one about the interband transition (due to the photon absorbtion) in electron-hole continuum region,
and the anormalous Rabi oscillation can emerges for the non-commuting Pauli matrices\cite{Kumar U} under the circularly polarized light
with complex vector potential.
For ARWA, the frequency obeys $\hbar\Omega\gg\varepsilon$ and $\hbar\Omega\gg\mathcal{A}$,
while for the rotating wave approximation (RWA) which is valid for the conventional Rabi oscillations
(in the absence of the pseudospin degree of freedom),
the detunning $\sim m_{D}^{\eta\sigma_{z}\tau_{z}}=\frac{1}{2}|2\hbar\Omega-2\varepsilon|\ll \varepsilon$ where $2\varepsilon$ is the energy of electron-hole pairs,
or due to the Zeeman shift\cite{Lin Y J}.
For the normal semiconductors as well as the silicene or graphene, the off-resonance light can effectively modifies the band structure 
as well as the quasienergy spectrum 
as shown in Fig.1(a),
while the resonance light is more likely to excites the atoms to a higher energy level.
Further, such modification is isotropic in the topological insulators but anisotropic in weyl semimetals due to the anomalous Bloch-Siegert shift.

For the first part of the slow oscillation term of the tight-binding Hamiltonian, we follow the expression of low-energy effective Hamiltonin in 
Ref.\cite{Ezawa M} as
\begin{equation} 
\begin{aligned}
H_{0}=\hbar v_{F}(\eta\tau_{x}k_{x}+\tau_{y}k_{y})+\eta\lambda_{{\rm SOC}}\tau_{z}\sigma_{z}+a\lambda_{R_{2}}\eta\tau_{z}(k_{y}\sigma_{x}-k_{x}\sigma_{y})
-\frac{\overline{\Delta}}{2}E_{\perp}\tau_{z}+\frac{\lambda_{R_{1}}}{2}(\eta\sigma_{y}\tau_{x}-\sigma_{x}\tau_{y}),
\end{aligned}
\end{equation}
where $E_{\perp}$ is the perpendicularly applied electric field, 
$\overline{\Delta}$ is the buckled distance in $z$-diraction between the upper sublattice and lower sublattice,
$\sigma_{z}$ and $\tau_{z}$ are the spin and sublattice (pseudospin) degrees of freedom, respectively.
$\eta=\pm 1$ for K and K' valley, respectively.
$\lambda_{SOC}=3.9$ meV is the strength of intrinsic spin-orbit coupling (SOC) and $\lambda_{R_{2}}=0.7$ meV is the intrinsic Rashba coupling
which is a next-nearest-neightbor (NNN) hopping term and breaks the lattice inversion symmetry.
$\lambda_{R_{1}}$ is the electric field-induced nearest-neighbor (NN) Rashba coupling which has been found that linear with the applied electric field
in our previous works\cite{Wu C H1,Wu C H2,Wu C H3,Wu C H4}, which as $\lambda_{R_{1}}=0.012E_{\perp}$.
For circularly polarized light with the electromagnetic vector potential has ${\bf A}(t)=A(\pm{\rm sin}\ \Omega t,{\rm cos}\ \Omega t)$,
where $\pm$ denotes the right and left polarization, respectively.
In this paper we use the radiation field with frequency $\Omega=3000$ THz$\approx 12$ eV,
which is certainly in the off-resonance region,
and thus the periodic gauge 
vector potential is ${\bf A}=\varepsilon_{0}/\Omega=0.078$ where $\varepsilon_{0}$ is the radiation field amplitude and we set $\hbar=e=1$ here, and
it satisfy ${\bf E}(t)=-\frac{1}{c}\partial_{t}{\bf A}(t)$.
Then the second part of the slow oscillation term is\cite{Kumar U}
\begin{equation} 
\begin{aligned}
H'=-\frac{e^{2}A^{2}a^{2}}{\hbar^{3}\Omega}[\eta\tau_{z}\hbar^{2}v_{F}^{2}-a^{2}\lambda_{R_{2}}^{2}\sigma_{z}
-a\lambda_{R_{2}}\hbar v_{F}(\eta\tau_{x}\sigma_{y}-\tau_{y}\sigma_{x})],
\end{aligned}
\end{equation}
which can also be written in the form of matrix as
\begin{equation} 
\begin{aligned}
H'_{K}=
\begin{pmatrix}[1.5] ia\lambda_{R_{2}}(A_{x}+iA_{y})&\hbar v_{F}(A_{x}-iA_{y})\\
\hbar v_{F}(A_{x}+iA_{y})&ia\lambda_{R_{2}}(A_{x}-iA_{y})
\end{pmatrix},\\
H'_{K'}=
\begin{pmatrix}[1.5] -ia\lambda_{R_{2}}(A_{x}+iA_{y})&-\hbar v_{F}(A_{x}+iA_{y})\\
-\hbar v_{F}(A_{x}-iA_{y})&-ia\lambda_{R_{2}}(A_{x}-iA_{y})
\end{pmatrix}.
\end{aligned}
\end{equation}
Through this, we can also find that
in the case of time-reversal-invariant (TRI; i.e., in the absence of Rashba coupling), $H'_{K}=H_{K'}^{'*}$.
However, that can be possible only for the linear-polarization case which with much smaller frequency,
since the circularly polarized light will breaks the TRI.
The fast oscillation term can be decribed as\cite{Ezawa M,Kitagawa T}
\begin{equation} 
\begin{aligned}
V=\frac{[V_{-1},V_{+1}]}{\Omega}+O(\frac{A^{4}}{\Omega^{2}})\approx\frac{\hbar^{2}v_{F}^{2}e^{2}a^{2}A^{2}}{\hbar^{2}\Omega}\tau_{z}\sigma_{z},
\end{aligned}
\end{equation}
where $V_{\mp 1}=\frac{1}{T}\int^{T}_{0}e^{\mp it\Omega}H(t)$, the Hubbard Hamiltonian here is $H(t)=t\sum_{ij,s}e^{i\phi n}c^{\dag}_{i,s}c_{j,s}$
with $n=|i-j|$, $\phi={\bf A}(t)/\phi_{0}$ where $\phi_{0}=h/e$ is the flux quanta. The on-site Hubbard term is ignored here.
Then the Floquet Hamiltonian can be written as $H_{F}=H_{0}+H'+V$,
and its periodicity $H_{F}(t+T)=H_{F}(t)$ can be immediately obtained through the periodicity Floquet states.
Note that the $\pm 1$ here denotes the harmonic order (or discrete Fourier index), 
For the high-order harmonic generation, the 
temporal behavior in off-resonance region can be described by the time-dependent Schr\"{o}dinger equation (TDSE):
\begin{equation} 
\begin{aligned}
i\frac{\partial}{\partial t}\psi(t)=H_{F}\psi(t),
\end{aligned}
\end{equation}
and the time propagation can be obatined by using the time-dependent generalized pseudospectral method\cite{Hermann M R,Chou Y} as
\begin{equation} 
\begin{aligned}
\psi(t)=e^{-i(H_{0}+H')t}\psi(0)
\end{aligned}
\end{equation}
which is approximated by the second-order split-operator technique 
\begin{equation} 
\begin{aligned}
\psi(t)=e^{-i(H_{0}+H')\frac{t}{2}}e^{-iV(r,\theta,\frac{t}{2})}e^{-i(H_{0}+H')\frac{t}{2}}\psi(0)+O(t^{3}),
\end{aligned}
\end{equation}
where the last term origin from the commutation errors.
From the above approximated form, we can obtain
\begin{equation} 
\begin{aligned}
\psi(t)=[{\rm cos}\ ((H_{0}+H')t)-i\ {\rm sin}\ ((H_{0}+H')t)]\psi(0),
\end{aligned}
\end{equation}
this result is agree with the Ref.\cite{Kumar U} base on the ideal that the Floquet Hamiltonian equals the half of the anomalous Rabi coupling 
by the off-resonance technique.

The solutions of the TDSE can be obtained by the high-order ($n$th) harmonic Floquet technique as 
$|\psi(t)\rangle=\sum_{n}e^{-i\varepsilon t/\hbar}e^{in\Omega t}|u_{n}\rangle$ where $|u_{n}\rangle$ is the $n$th Bloch state
and the corresponding Dirac quasienergy spectrum 
\begin{equation} 
\begin{aligned}
\varepsilon=\tau_{z}\Gamma_{R}-\frac{1}{2}\eta\alpha\hbar\Omega,
\end{aligned}
\end{equation}
where $\Gamma_{R}$ is the generalized Rabi frequency, and here the parameter $\alpha=1,0,1,0\cdot\cdot\cdot$ 
corresponds to the momentum $k=0,1,2,3\cdot\cdot\cdot$, respectively, in the normal case.
But it's $\alpha=0,1,0,1\cdot\cdot\cdot$ corresponds to the momentum $k=0,1,2,3\cdot\cdot\cdot$ in the Dirac-point anticrossing case\cite{Perez-Piskunow P M},
which can be seen typically in the quantum anomalous Hall phase\cite{Wu C H2}.
Here the momentum $k$ is near the Dirac-point ($k=0$ in Dirac-point).
Note that here we here we restrict the Floquet zone within the energy range $(-\hbar\Omega/2,\hbar\Omega/2]$ as done in Ref.\cite{Perez-Piskunow P M}
and the degeneracy between the hamonic order and the energy spectrum is lifted by the circularly polarized light
which can also be seem by the winding number.
The above energy range is base on the assumation of a second-order process of 
emit $n$ photons and then absorpt $m$ photons, $\alpha=0$ corresponds to the $|n-m|=0$ case, while $\alpha=1$ corresponds to the $|n-m|=1$ case.
While in the near- or below-critical frequency region, the first-order process happen even in the semiclassical ($n\gg 1$) situation as:
after the atoms firstly tunnels the barrier potential formed by the radiation driving, the combined barrier potential is hard fot the atoms to
escape again, thus then the atoms will absorbs serveral photons and then return to the ground state quickly which form the so-called "multi-photon" and
"multi-rescattering" process\cite{Li P C}.

The dynamical generator about the radiation driving can also be obtained as\cite{Kumar V}
\begin{equation} 
\begin{aligned}
H=H_{0}+\hbar v_{F}\mathcal{A}[e^{in\Omega t}c_{A}^{\dag}c_{B}b^{\dag}+e^{-in\Omega t}c_{A}c_{B}^{\dag}b],
\end{aligned}
\end{equation}
where the dimensionless intensity $\mathcal{A}=eAa/\hbar$ is in a form similar to the Bloch frequency, 
and it's estimated as 0.3 here.
$c_{A}^{\dag}$ and $c_{B}$ are the creation operator and annihilation operator of the sublattice $A$ and $B$, respectively,
and $[b,b^{\dag}]=1$ denotes the photon operator.
The Floquet eigenstates are 
\begin{equation} 
\begin{aligned}
|\psi(t)\rangle^{K}_{\tau_{z},\sigma_{z}}=\frac{e^{-i\varepsilon t}}{\sqrt{2}}\begin{pmatrix}e^{-in\Omega t}\sqrt{1+\frac{\tau_{z}m_{D}^{+\sigma_{z}\tau_{z}}}{\Gamma_{R}}}\\
\tau_{z}\sqrt{1-\frac{\tau_{z}m_{D}^{+\sigma_{z}\tau_{z}}}{\Gamma_{R}}}
\end{pmatrix}
\end{aligned}
\end{equation}
for $K$ valley, and
\begin{equation} 
\begin{aligned}
|\psi(t)\rangle^{K'}_{\tau_{z},\sigma_{z}}=\frac{e^{-i\varepsilon t}}{\sqrt{2}}\begin{pmatrix}e^{in\Omega t}\sqrt{1+\frac{\tau_{z}m_{D}^{-\sigma_{z}\tau_{z}}}{\Gamma_{R}}}\\
\tau_{z}\sqrt{1-\frac{\tau_{z}m_{D}^{-\sigma_{z}\tau_{z}}}{\Gamma_{R}}}
\end{pmatrix}
\end{aligned}
\end{equation}
for $K'$ valley.

The generalized Rabi frequency for the monolayer silicene is
\begin{equation} 
\begin{aligned}
\Gamma_{R}=\sqrt{\hbar^{2}v_{F}^{2}({\bf k}+{\mathcal{A}})^{2}+(m_{D}^{\eta\sigma_{z}\tau_{z}})^{2}},
\end{aligned}
\end{equation}
and the momentum-independent Dirac-mass here is 
\begin{equation} 
\begin{aligned}
m_{D}^{\eta\sigma_{z}\tau_{z}}=|\eta\lambda_{{\rm SOC}}s_{z}\tau_{z}-\frac{\overline{\Delta}}{2}E_{\perp}\tau_{z}+Ms_{z}-\eta\hbar v_{F}^{2}\frac{\mathcal{A}}{\Omega}|,
\end{aligned}
\end{equation}
with 
\begin{equation} 
\begin{aligned}
\hbar v_{F}{\bf k}=\hbar v_{F}k\begin{pmatrix}0 &k_{x}-ik_{y}\\ k_{x}+ik_{y}& 0\end{pmatrix}
=\pm t|1+e^{ik_{x}}+e^{-ik_{y}}|,
\end{aligned}
\end{equation}
where $k_{x}+ik_{y}=\hbar v_{F}\vec{\sigma}_{AB}\cdot{\bf k}$ with the vector $\vec{\sigma}_{AB}=\sigma_{z}\cdot({\rm cos}(E_{ex}\Phi),{\rm sin}(E_{ex}\Phi),0)$
describes the adiabatic evolution with the variable $\Phi$ with the induced frustrations.
For pseudospin, it's usually takes $\Phi=\phi={\rm arctan}\frac{k_{y}}{k_{x}}$.
$E_{ex}$ is the exchange field which related to the chirality.
Here we comment that the exchange field for the out-of-plane polarization is much larger than the in-plane one.
Note that here we don't consider the $\lambda_{R_{1}}$ and $\lambda_{R_{2}}$ term in the above Dirac-mass,
if they are contained, the Dirac-mass becomes
\begin{equation} 
\begin{aligned}
m_{D}^{'\eta\sigma_{z}}=|\eta\sqrt{\lambda_{{\rm SOC}}^{2}+a^{2}\lambda_{R_{2}}^{2}}s_{z}\tau_{z}
-\frac{\overline{\Delta}}{2}E_{\perp}\tau_{z}+Ms_{z}-\eta\hbar v_{F}^{2}\frac{\mathcal{A}}{\Omega}+\lambda_{R_{1}}/2|.
\end{aligned}
\end{equation}
While for the bilayer silicene, the Rabi frequency is 
\begin{equation} 
\begin{aligned}
\Gamma_{R}=\sqrt{\left[\frac{\hbar^{2}({\bf k}+{\mathcal{A}})^{2}}{2m^{*}}\right]^{2}+(m_{D}^{\eta\sigma_{z}\tau_{z}})^{2}},
\end{aligned}
\end{equation}
with 
\begin{equation} 
\begin{aligned}
\frac{\hbar^{2}{\bf k}^{2}}{2m^{*}}=\frac{{\bf k}^{2}t^{2}}{t'}
=\frac{\hbar^{2}k^{2}}{2m^{*}}\begin{pmatrix}0 &(k_{x}-ik_{y})^{2}\\ (k_{x}+ik_{y})^{2}& 0\end{pmatrix}=
\pm\frac{t'}{2}\pm\sqrt{(\frac{t'}{2})^{2}+t^{2} |1+e^{ik_{x}}+e^{-ik_{y}}|^{2}}  
\end{aligned}
\end{equation}
where $t'$ is the interlayer hopping,
that can be easily deduced by $m^{*}=\hbar^{2}t'/(2t^{2})$ and $\hbar v_{F}=\frac{\sqrt{3}}{2}at$.
The above expression results in the four band structure in the spin degenerate case for bilayer silicene.
It's obviously that the Rabi frequency of monolayer is linearly ${\bf k}$-dependent while the bilayer one is the quadratic-dispersion,
that's similar to their energy spectrum.
For the monolayer silicene and bilayer silicene, 
the difference in the phase factor as well as the exchange field $E_{ex}$ 
origin from the diffferece of Berry phase as well as the rotation of pseudospin between the monolayer silicene and bilayer silicene,
which are\cite{Park C H}
\begin{equation} 
\begin{aligned}
\gamma_{mono}=-i\lim_{n\rightarrow\infty}\sum^{n-1}_{i=0}{\rm log}(\frac{1+e^{i\phi}}{2})=\pi,\\
\gamma_{bi}  =-i\lim_{n\rightarrow\infty}\sum^{n-1}_{i=0}{\rm log}(\frac{1+e^{2i\phi}}{2})=2\pi,\\
\end{aligned}
\end{equation}
where $\phi={\rm arctan}\frac{{\bf k}_{i+1}-{\bf k}_{i}}{{\bf k}_{i}}$.
The anomalous Rabi oscillation frequency is
\begin{equation} 
\begin{aligned}
\Gamma_{AR}=\sqrt{\hbar^{2}v_{F}^{2}({\bf k}+{\mathcal{A}})^{2}+(m_{D}^{\eta\sigma_{z}\tau_{z}}\pm 
\frac{e^{2}\mathcal{A}^{2}v_{F}^{2}}{c^{2}}\vec{\sigma}_{AB}\frac{1}{\Omega})^{2}},
\end{aligned}
\end{equation}
and the corresponding eigenvalue (i.e., the quasienergy spectrum) is $\varepsilon_{AR}=\tau_{z}\Gamma_{AR}-\frac{1}{2}\eta\alpha\hbar\Omega$.
The anomalous Rabi oscillation frequency is unlike the conventional Rabi frequency\cite{Kumar U},
it's related to the Chern number: when the Chern number of silicene (or for other topological insulator system) is zero, then the induced (by 
anomalous Rabi oscillation frequency) mass is nonzero for this trivial system and thus with the gapped edge states;
when the Chern number is nonzero, the induced mass is zero and corresponds to the non-trivial system with the gapless edge states.

For bilayer silicene, we consider the AB-stacked bilayer silicene in this article which with the NN interlayer hopping as $t=2$ eV
and we ignore the NNN interlayer hopping.
The interlayer SOC is estimated as 0.5 meV here\cite{Ezawa M2} and since the 
trigonal warping term between two layers has a non-negligible impact when apply the light in terahert range\cite{Morell E S},
we set the trigonal warping hopping parameter as $t_{w}=0.16$ eV.
Then the low-energy Dirac effective model can be written in a matrix form
\begin{equation} 
\begin{aligned}
H_{bi}=\begin{pmatrix}m_{D}^{\eta ++}-\eta\hbar v_{F}^{2}\frac{\mathcal{A}^{2}\Omega}{t^{'2}}&\hbar v_{w}(k_{x}+ik_{y})&0&\hbar v_{F}(k_{x}-ik_{y})\\
 \hbar v_{w}(k_{x}-ik_{y})    &m_{D}^{\eta +-}+\eta\hbar v_{F}^{2}\frac{\mathcal{A}^{2}\Omega}{t^{'2}}&\hbar v_{F}(k_{x}+ik_{y})&0\\
 0&\hbar v_{F}(k_{x}-ik_{y})   &m_{D}^{\eta -+}&\eta t'\\        
 \hbar v_{F}(k_{x}+ik_{y})   &0&\eta t'&m_{D}^{\eta --}\\  
 \end{pmatrix},
\end{aligned}
\end{equation}
where $v_{w}=\sqrt{3}at_{w}/2\hbar$ is the velocity associates with the trigonal warping.
Because of the existence of trigonal warping term in above Hamiltonian, the valley symmetry is broken which may leads to the single-Dirac-cone state,
that implys the light in a finite intensity has the same effect with the out-of-plane antiferromagnetic exchange field.
The asymmetry effect on the single valley band structure can be seen in the Fig.1(b),
where we consider the NN Rashba coupling but ignore the exchange field.

\section{Dynamical polarization}
The radiational driving also leads to the dramatic oscillation of the pseudospin polarization $\tau_{z}=|c^{\dag}_{A}c_{A}-c^{\dag}_{B}c_{B}|$ 
unlike the original one
which is $\tau'_{z}=m_{D}^{\eta\sigma_{z}\tau_{z}}/\sqrt{\hbar^{2}v_{F}^{2}{\bf k}^{2}+(m_{D}^{\eta\sigma_{z}\tau_{z}})^{2}}$.
The resulting pseudospin polarization is 
\begin{equation} 
\begin{aligned}
\tau_{z}(t)=&\frac{2\hbar v_{F}\mathcal{A}}{\Gamma_{AR}}\frac{\hbar^{2}v_{F}^{2}{\bf k}^{2}}{\Gamma_{AR}}{\rm sin}(\Gamma_{AR}t)
(\frac{m_{D}^{\eta\sigma_{z}\tau_{z}}}{\Gamma_{AR}}{\rm sin}(\Gamma_{AR}t){\rm cos}\varphi-{\rm cos}(\Gamma_{AR}t){\rm sin}\varphi)\\
&+\frac{m_{D}^{\eta\sigma_{z}\tau_{z}}}{\Gamma_{AR}}(1-\frac{2\hbar^{2}v_{F}^{2}\mathcal{A}^{2}}{\Gamma_{AR}^{2}}{\rm sin}^{2}(\Gamma_{AR}t)),
\end{aligned}
\end{equation}
where $\varphi\in(0,2\pi]$ over the Bloch sphere and can be estimated as $\varphi={\rm arctan}(\varepsilon_{j}/\varepsilon_{i})$
for $\varepsilon=\varepsilon_{i}\vec{i}+\varepsilon_{j}\vec{j}$.

While for the valley polarization at finite temperature,
it's proportional to the term $\sum_{i}(f_{|K+{\bf k}_{i}|}+f_{|K'+{\bf k}_{i}|})$ in unit cell where $f$ is the Fermi-Dirac distribution function. 
Through the expression of $\varepsilon_{AR}$ obtained above,
we can write the valley polarization as
\begin{equation} 
\begin{aligned}
\eta(t)=2\beta\frac{36\hbar v_{F}^{2}}{a^{4}}f(\hbar\Omega)\left[\frac{{\rm sin}^{2}[(\Gamma_{AR}-\frac{\hbar\Omega}{2})t]}{(\Gamma_{AR}-\frac{\hbar\Omega}{2})^{2}}
+\frac{{\rm sin}^{2}[(\Gamma_{AR}+\frac{\hbar\Omega}{2})t]}{(\Gamma_{AR}+\frac{\hbar\Omega}{2})^{2}}\right],
\end{aligned}
\end{equation}
where the factor $\beta\approx$ 21.5 
incorporate the quasienergy- and momentum-dependent normalization\cite{Aghel F} factor which is to constrain the valley polarization into $[-1,1]$
and obtain zero time-average value.

For the out-of-plane spin polarization, 
through the above Floquet technique, the wave function of the dressed state at zeroth discrete Fourier index ($n=0$) and at Dirac-point ($\alpha=0$) 
can be written as
\begin{equation} 
\begin{aligned}
|\psi(t)\rangle=\frac{\hbar^{2}v_{F}^{2}({\bf k}+\mathcal{A})^{2}}{\Gamma_{AR}^{2}}
\frac{\hbar^{2}v_{F}^{2}({\bf k}+\mathcal{A})^{2}-\Gamma_{AR}^{2}}{i\Gamma_{AR}m_{D}^{\eta\sigma_{z}\tau_{z}}}e^{-i\varepsilon_{AR} t}
\end{aligned}
\end{equation}
Thus the out-of-plane spin polarization reads
\begin{equation} 
\begin{aligned}
\sigma_{z}=4\frac{\hbar^{2}v_{F}^{2}({\bf k}+\mathcal{A})^{2}}{\Gamma_{AR}^{2}}
\frac{\hbar^{2}v_{F}^{2}({\bf k}+\mathcal{A})^{2}-\Gamma_{AR}^{2}}{i\Gamma_{AR}m_{D}^{\eta\sigma_{z}\tau_{z}}}{\rm cos}(\varepsilon_{AR}E_{ex}t),
\end{aligned}
\end{equation}
where the factor 4 denotes the valley and pseudospin degrees of freedom,
and the term ${\rm cos}(E_{ex}t)$ is the spin-exchange-induced frustration to the hexgonal lattice system of monolayer silicene,
and $E_{{\rm ex}}$ is the intralayer exchange energy induced by the $z$-direction spin which with preserved total angular momentum.
The total angular momentum is thus commute with the static effective Hamiltonian $H_{0}$.
$\sigma_{z}$ is a good quantum number with conserved $z$-direction spin when the Rashab-coupling is ignored.
Since for monolayer silicene (as well as the two-dimension electron gas (2DEG) or graphene), the exchange energy is half of the bilayer one,
for simplification, we set $E_{{\rm ex}}=1$ for monolayer silicene and $E_{{\rm ex}}=2$ for bilayer silicene in our computation.

\subsection{Simulation results and discussion}

The pseudospin polarization against the momentum under different electric field is presented in Fig.2,
where the angle over the Bloch sphere $\phi$ is setted as $\pi/4$.
The oscillation of pseudospin polarization exhibits the beating structure, 
which will vanish for the pseudospinless model (with constant amplitude).
In Fig.2,
a abnormal hump near the ${\bf k}=0$ point (Dirac-point) is arised with the increasing electric field.
We also find that the oscillation of the pseudospin polarization is periodic and follows a ${\bf k}=0.6$ cycle
(see the enlarged view in the right-side of Fig.2(a)-(c))
in the flat region (far away from ${\bf k}=0$).
The pseudospin polarization here shows the collapse-and-revival behavior 
(or the construction-destruction interference) with very short dephasing time (relaxation time)
which is related to the pulse duration.
By comparing to the long-time panel (Fig.1(d)),
we found that the electric field affects little to the oscillation periodic and the collapse-and-revival pattern,
while the time affects the oscillation periodic and the beating more.
Such collapse-and-revival behavior also be found in the Rabi oscillation with the non-constant amplitude, e.g.,
the Rabi oscillation can been seen in the Bose-Einstein condensate trapped in a harmonic potential which acts like a Zeeman field
when without applying the driving field\cite{Zhang Y},
and it can be solved by the momentum dipole method for the oscillations of center-of-mass (COM) with $\tau_{z}=1$ component and $\tau_{z}=-1$ component.
The radiational driving leads to the dramatic oscillation of the pseudospin polarization configuration near the Dirac-cone
and breaks the valley symmetry in momentum space\cite{Min H},
which is different from the original configuration (hedgehog-type meron in the momentum space).
The orientation of pseudospin in the bilayer silicene changes rapidly due to the transfer of charges between two layers\cite{Min H},
with the broken inversion symmetry between two layers due to the valley-asymmetry induced by the circularly polarized light
(or caused by the potential difference between two layers).

For the electron-hole symmetry case, which happen when the chemical potential is zero and thus with two equivalent sublattices,
the nonrelativistic effect will emerges in the Dirac-point which with zero Rashba coupling,
and the spin and pseudospin degrees of freedom are decoupled by this effect.
The photocurrent is observable at only half of the unit cell in such case.
In this article, we always set the chemical potential as 0.2 eV,
which slightly breaks the electron-hole symmetry and thus gives rise the relativistic effect and lifts the degeneracy between the spin and pseudospin
which can be clearly be seen in the photoemission spectrum (quadratic dispersion) with the in-plane spin polarized quasiparticles under bias voltage\cite{Kuemmeth F}.
In this case, the giant spin-orbit splitting can be observed with the metallic substrate\cite{Marchenko D} due to the strong spin-orbit interaction
and the symmetry-breaking due to the substrate-induced potential-difference.
However, although the existence of finite chemical potential, the nonrelativistic effect is still exist in the $\Gamma$-point (the center of 
first Brillouin zone), whose momentum-independent isotropy 
(with nearly circle shape in the isoenergy surface) won't be affected by the photoemission anisotrpy.
In Fig.3, we show the rate of change of the pseudospin polarization $(d/dt)\tau_{z}$ against time.
The electric field affects less to the polarization than the momentum (or the distance to Dirac-point).
It shows that, the period of oscillation and the beating increase with the increasing of momentum.

In Fig.4, we present the results of the valley polarization near Dirac-point (we set ${\bf k}$=2 here) under different strengths of electric field,
where the temperature is setted as $T=20$ K.
From Eq.(23), we can know that the temperature-dependent Fermi-Dirac distribution function $f(\hbar\Omega)$ determines the amplitude of the oscillation
and it also affects the collapse-and-revival pattern as well as the beating\cite{Fahandezh Saadi M}.
Unlike the pseudospin polarization, the electric field enhance the beating obviously in low-electric field range;
however, for the large electric field ($E_{\perp}\gg 20$ eV), the evolution of beating is irregular with the electric field.
The oscillation of valley polarization is related to the electron-phonon scattering due to the photoexcitation
which has a relaxation time in picosecond range\cite{Kumar S} and larger than that of the electron-eletron scattering.
Fig.5 shows the temporal behavior momentum-dependent out-of-plane spin polarization near the Dirac-point (${\bf k}=2$ and with $\alpha=0$ at $n=1$ Fourier order) of monolayer silicene,
where we assume it's begin with a nonzero spin polarization
and the Rashba coupling is ignored here.
The exchange field between the driving field and the two spin-component is setted as 1 for the monolayer silicene, 
which is half of that of the bilayer silicene.
The amplitude of the oscillation of spin polarization is increase with the increasing electric field at low electric field regime
(the amplitude behaviors irregularly with the applied electric field for the case of $E_{\perp}\gg\Omega$).

\section{Motion of the center of mass}
As we mentioned in above, the collapse-and-revival pattern can also be found in
the oscillations of center of mass (COM) under the laser effect which can be detected by the momentum dipole method 
with components $\tau_{z}=1$ and $\tau_{z}=-1$.
At the COM, the wave vector can be expressed as ${\bf q}_{COM}={\bf k}(\tau_{z})-{\bf k}(-\tau_{z})$ for the monolayer silicene\cite{Zhou X}.
Although the existence of the large intrinsic SOC in silicene, the COM won't couples the spin and the motion of COM\cite{Lin Y J},
however, the motion of COM is affected by the strength of SOC as well as the chirp of laser, which we setted as $\kappa=1$ ps$^{2}$ here.
We assume the initial coordinates of COM for both the two pseudospin components are $x_{0}=2$,
and the width of the dynamical wave package is in a effective characteristic scale (effective Bohr radius) $R=0.3r_{Si-Si}=0.648$ \AA\,
while for the phase difference of dipole oscillation between the two components,
is setted $\Delta\phi=0.9\pi$ here and it's equals tp $\pi$ for vanishing SOC.
in addition, we ignore the transverse frequency here and consider only the longitudinal Rabi freuency,
then the expression of COM in the off-resonance regime satisfies the following approximation relation
\begin{equation} 
\begin{aligned}
x(t)\approx &\tau_{z}(\lambda_{SOC}-\kappa)-\tau_{z}\frac{2\sqrt{2}}{-2/R^{2}}
{\rm exp}(\frac{4\kappa^{2}-4x_{0}^{2}/R^{4}}{-4x_{0}/R^{2}}-\frac{x_{0}^{2}}{R^{2}})\frac{1}{\sqrt{2}}\Gamma_{AR}\\
&[2\kappa{\rm cos}(\frac{4\kappa x_{0}/R^{2}}{-2/R^{2}}t)-(\frac{2x_{0}}{R^{2}}-\frac{2}{R^{2}})
{\rm sin}(\frac{4\kappa x_{0}/R^{2}}{-2/R^{2}}t)]
\end{aligned}
\end{equation}
which can be deduced from the approximated Gross-Pitaevskii equations under the non-resonance circularly polarized light: 
$i\frac{\partial\psi}{\partial t}\propto \Gamma_{AR}\psi$.
It's obviously that the expression of COM is a superposition of the cosinoidal function and sinusoidal function except for the $x_{0}=1$.
Fig.6 shows the result of motion of COM.
We can see that the oscillations of the up-pseudospin and down-pseudospin are in opposite directions,
and the beating (collapse-and-revivals pattern) vanish in the oscillation of COM,
that's due to the some value of widths $R$ and initial coordinates between two pseudospin components setted by us.
In fact, except the difference of $R$ and initial coordinates between two pseudospin components,
the beating of motion of COM is also related to the interaction strength between two pseudospins.

\section{Conclusion}
In this paper, we explore the dynamic under the off-resonance circularly polarized light as well as
the dynamical polarization of the pseudospin, valley, and spin degrees of freedom,
the expressions of these polarizations under the radiation off-resonance driving field are presented,
and we found the collapse-and-revivals pattern is exist in these degrees of freedom under a certain condition,
and its period is related to the quantum optics.
The off-resonance laser can also induce the topological phase transition by manipulating the energy band structure,
rather than excite the atoms to the high quantum-number states like the resonance light.
It's also found than the momentum-dependence of the spin-polarization may increase when depositing the silicene on a Au/Ni(111) substrate\cite{Rashba E I}.
Except for the radiation driving field,
the exchange biased system can also effectively affects the interfacial spin configuration\cite{Shiratsuchi Y} in experiments.
The collapse-and-revivals pattern can be found in the 
the Rabi oscillation which exist in the Bose-Einstein
condensate trapped in a harmonic potential and acts like a Zeeman field when without apply
the driving field,
or by the periodic energy-exchange (by the exchange field which related to the chirality)  
between the radiation driving field and the two-component dynamical polarization when apply the circularly polarized light in off-resonance regime.
Additionally, due to the exist of scattering by the charged impurities,
the quasiparticle oscillation will damp until reaches the equilibrium state\cite{Dóra B}.
The motion of COM is explored, which would also exhibits the collapse-and-revivals pattern for unequal initial coordinate $x_{0}$ and
and the width of the dynamical wave package $R$ but with zero beating for equal $x_{0}$ and $R$.
The oscillations of the up-pseudospin and down-pseudospin are in opposite directions,
and they are related to the anormalous Rabi frequency just like the above-mentioned dynamical polarizations
due to the effect of circularly polarized light.
Our results can also be applied to the two-dimension low-energy Dirac models or the surface of the three-dimension topological insulators,
and even the weyl semimetal with the photoinduced topological phase transition\cite{Kumar U}.


\end{large}
\renewcommand\refname{References}

\clearpage
Fig.1
\begin{figure}[!ht]
   \centering
   \centering
   \begin{center}
     \includegraphics*[width=0.5\linewidth]{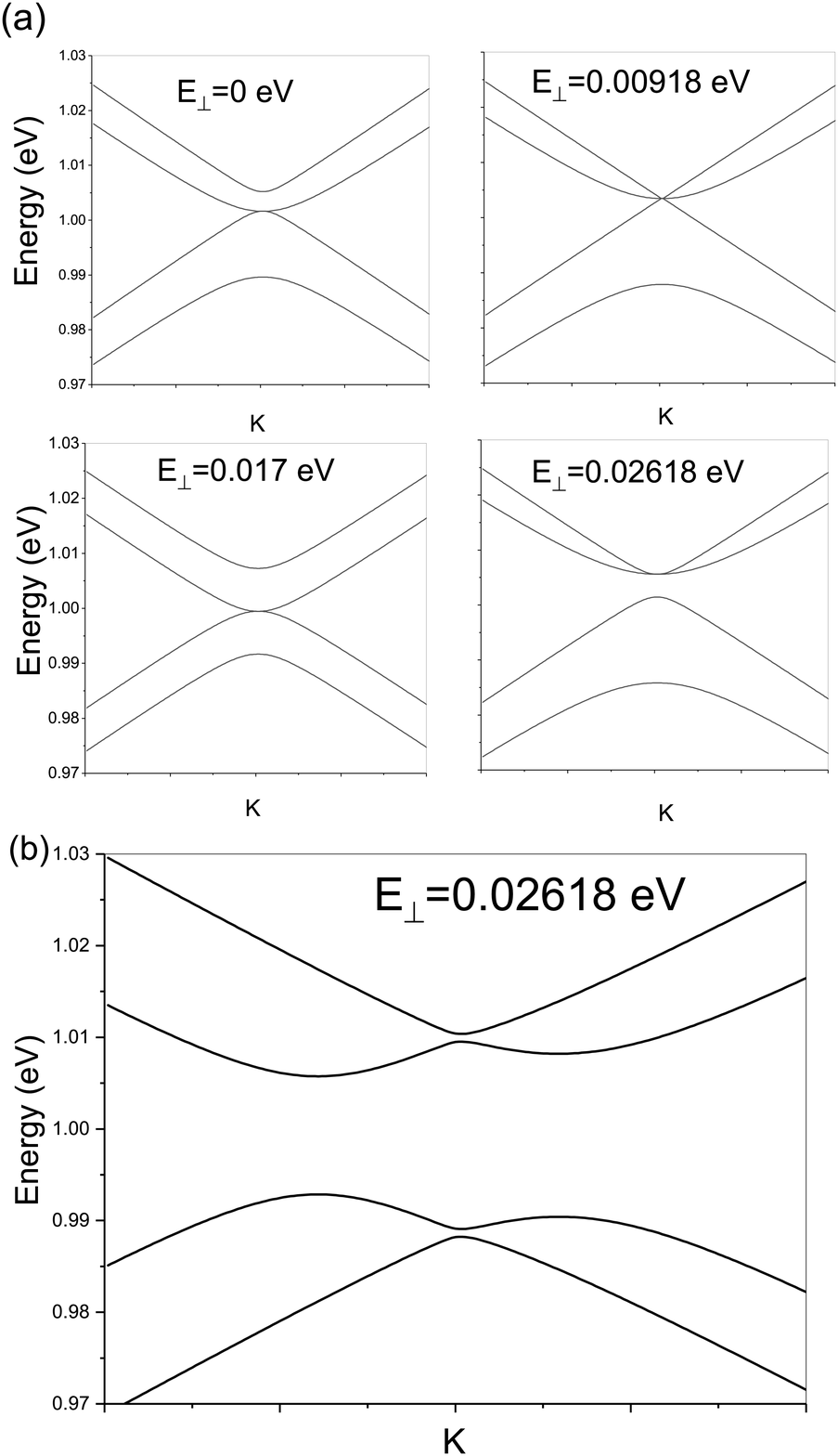}
\caption{(a) Band gap evolution at K valley for monolayer silicene under the electric field and radiational driving field.
The exchange field $M$ is setted as 0.0039 eV, and electric-field-dependent NN Rahsba coupling is considered here.
The electric field is setted as $E_{\perp}=0,\ 0.00918,\ 0.017,\ 0.02618$ eV as labeled in the plots.
(b) is for the bilayer silicene under electric field $E_{\perp}=0.02618$ eV which we don't consider the exchange field (thus the symmetry between conduction band and valence band is retain)
but consider the trigonal warping term as $t_{w}=0.16$ eV and the interlayer hopping $t'=2$ eV.
The valley asymmetry due to the trigonal warping term can be easily seen.
}
   \end{center}
\end{figure}
\clearpage
Fig.2
\begin{figure}[!ht]
   \centering
\subfigure{
\begin{minipage}[t]{0.4\textwidth}
\includegraphics[width=1\linewidth]{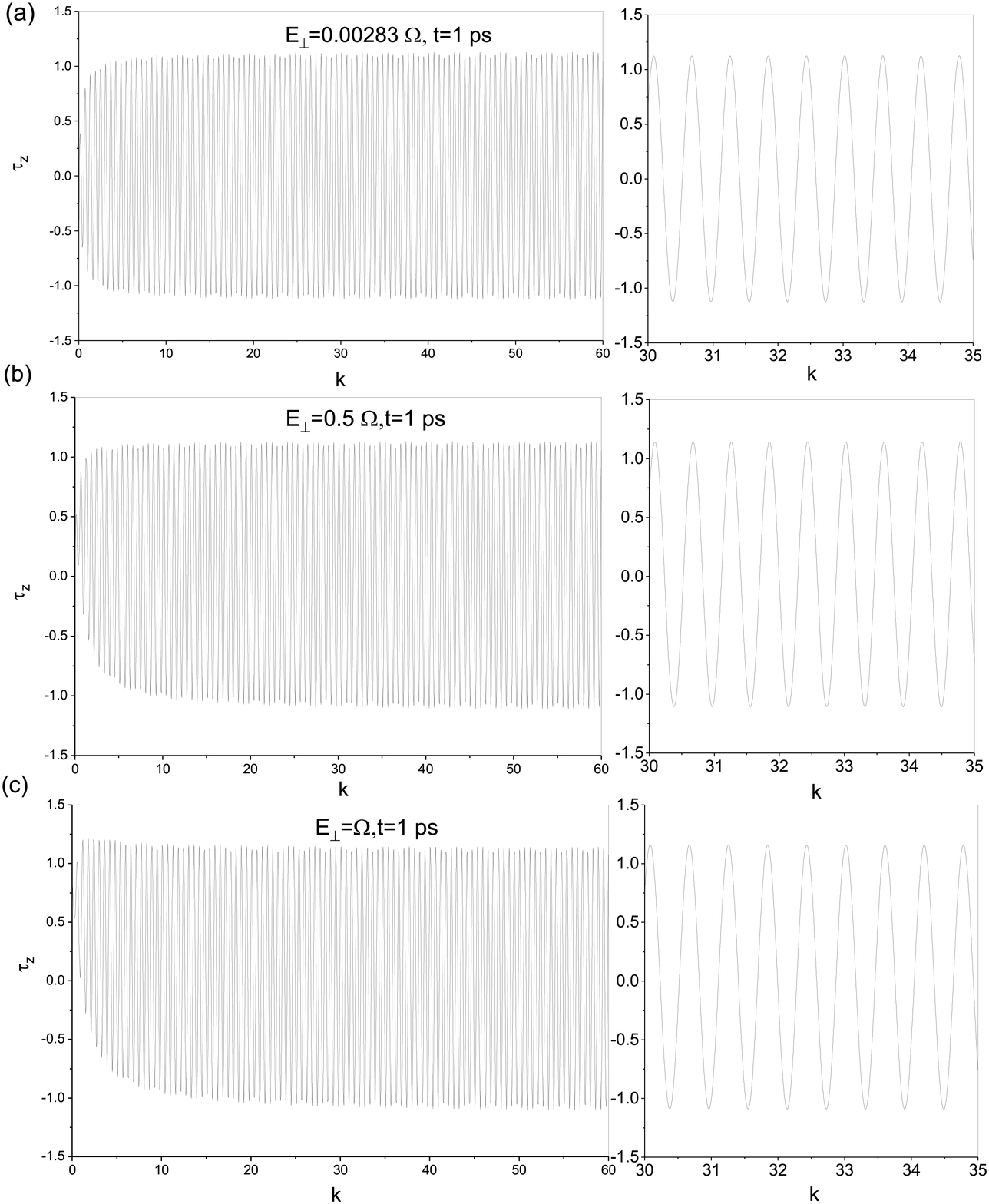}
\label{fig:side:b}
\end{minipage}
}\\
\subfigure{
\begin{minipage}[t]{0.4\textwidth}
\includegraphics[width=1\linewidth]{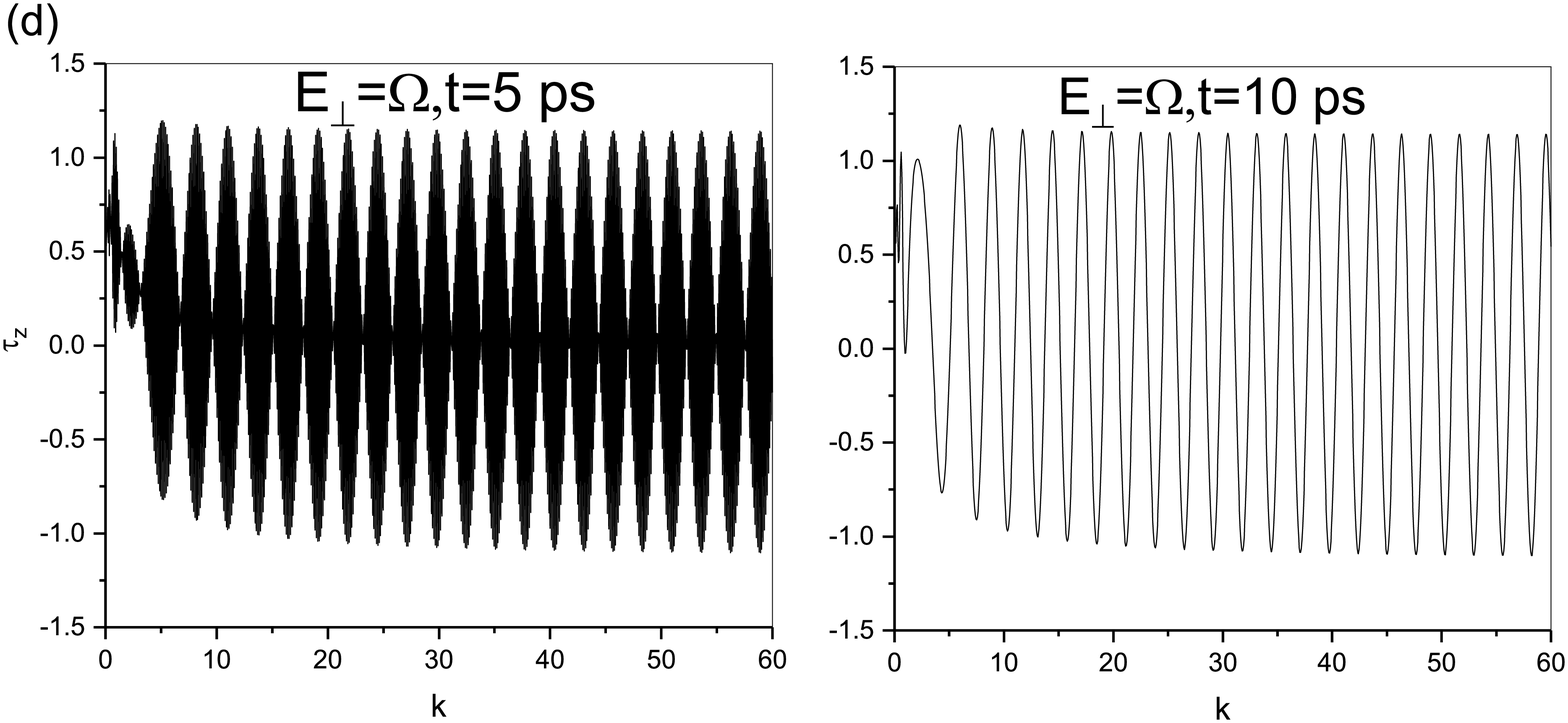}
\label{fig:side:a}
\end{minipage}
}
\caption{(a)-(c) Pseudospin polarization as a function of the momentum in a fixed time $t=1$ ps.
The Rashba coupling is ignored here. (d) The pseudospin polarization as a function of the momentum at the time $t=5$ ps and $t=10$ ps.
}
\end{figure}
\clearpage
Fig.3
\begin{figure}[!ht]
   \centering
   \begin{center}
     \includegraphics*[width=0.8\linewidth]{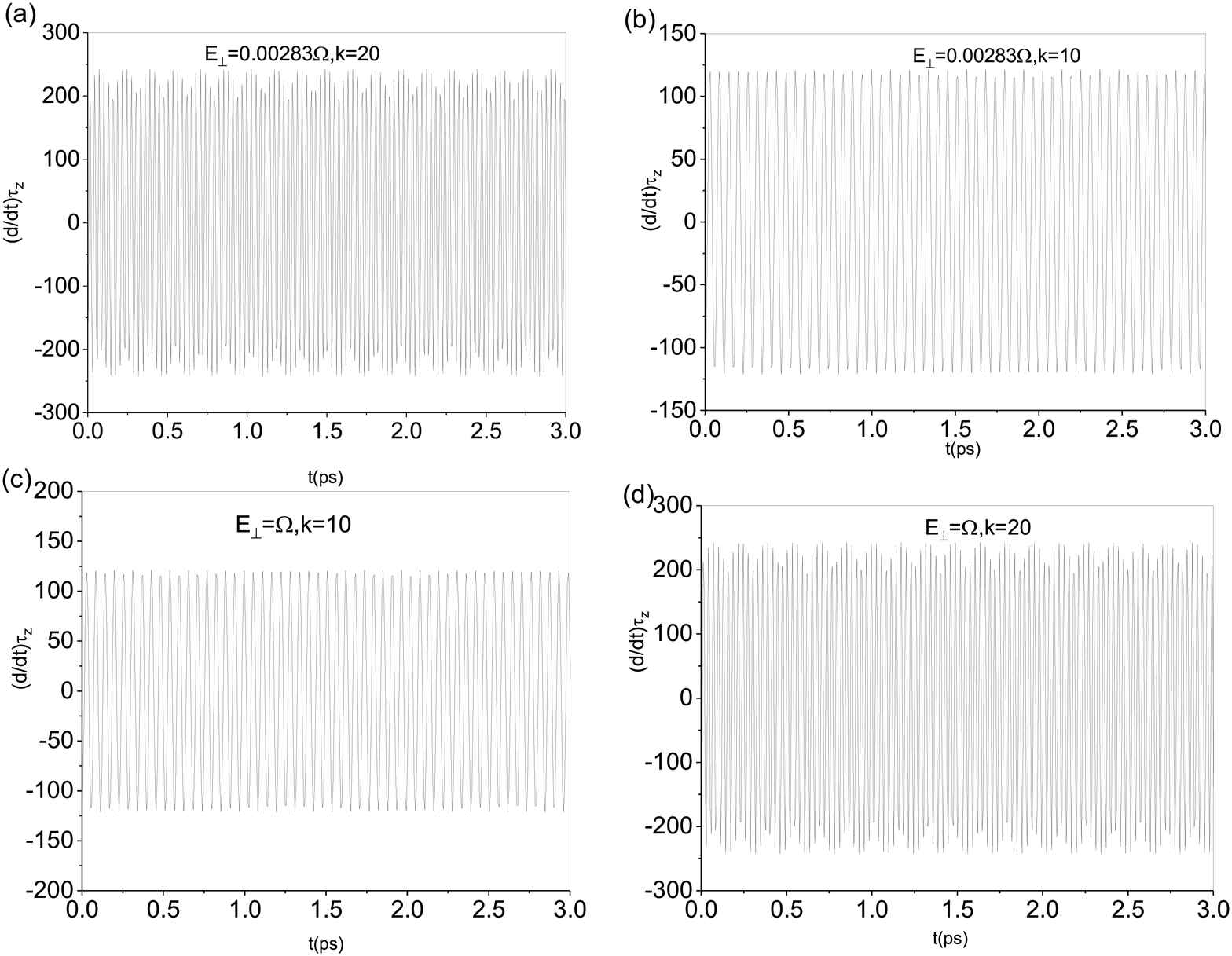}
\caption{Equation of motion of the pseudospin polarization against time (in unit of ps since the frequency of light is setted in the terahertz range)
at different electric field and momentum.
}
   \end{center}
\end{figure}
\clearpage
Fig.4
\begin{figure}[!ht]
   \centering
   \begin{center}
     \includegraphics*[width=0.8\linewidth]{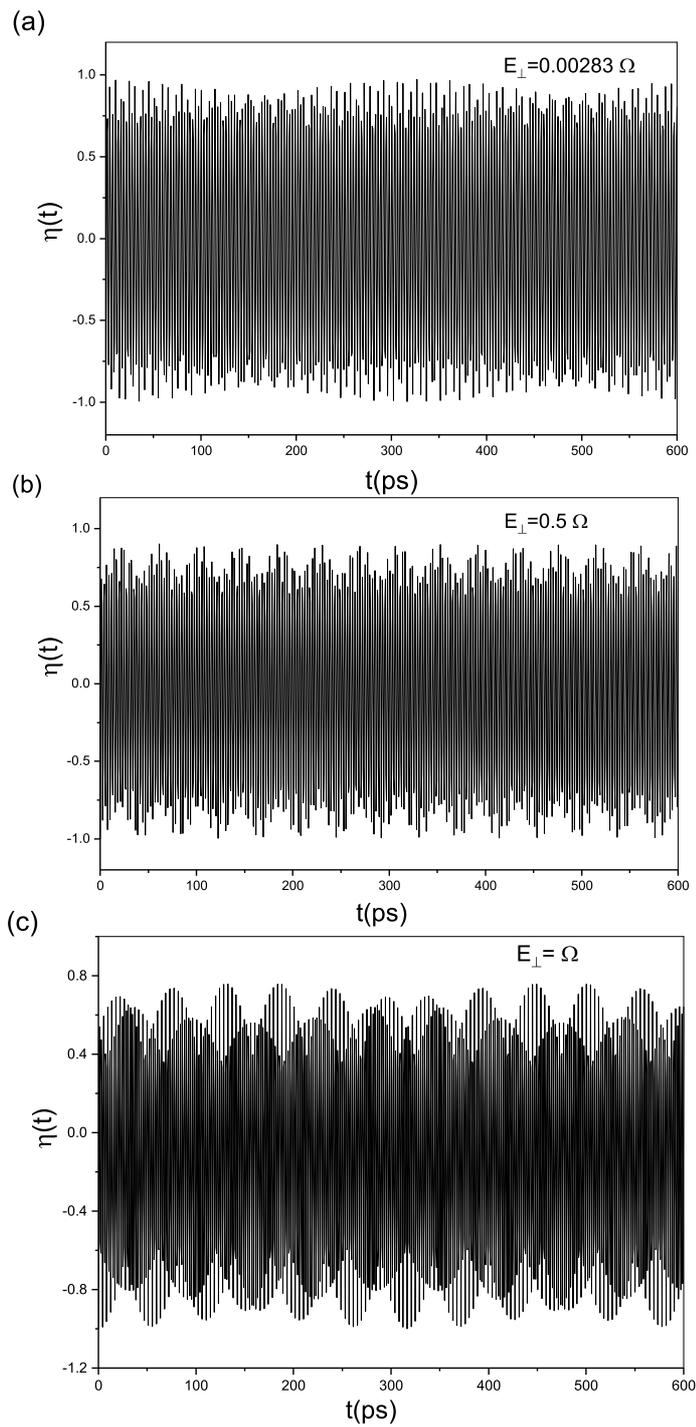}
\caption{Valley polarization as a function of time at momentum ${\bf k}=2$ under different strengths of electric field.
The temperature is setted as $T=20$ K.
The frequency of light used here is $3000$ THz$=7.5$ t.
The corresponding electric field are labeled in the plots.
}
   \end{center}
\end{figure}
\clearpage
Fig.5
\begin{figure}[!ht]
   \centering
\subfigure{
\begin{minipage}[t]{0.4\textwidth}
\includegraphics[width=1\linewidth]{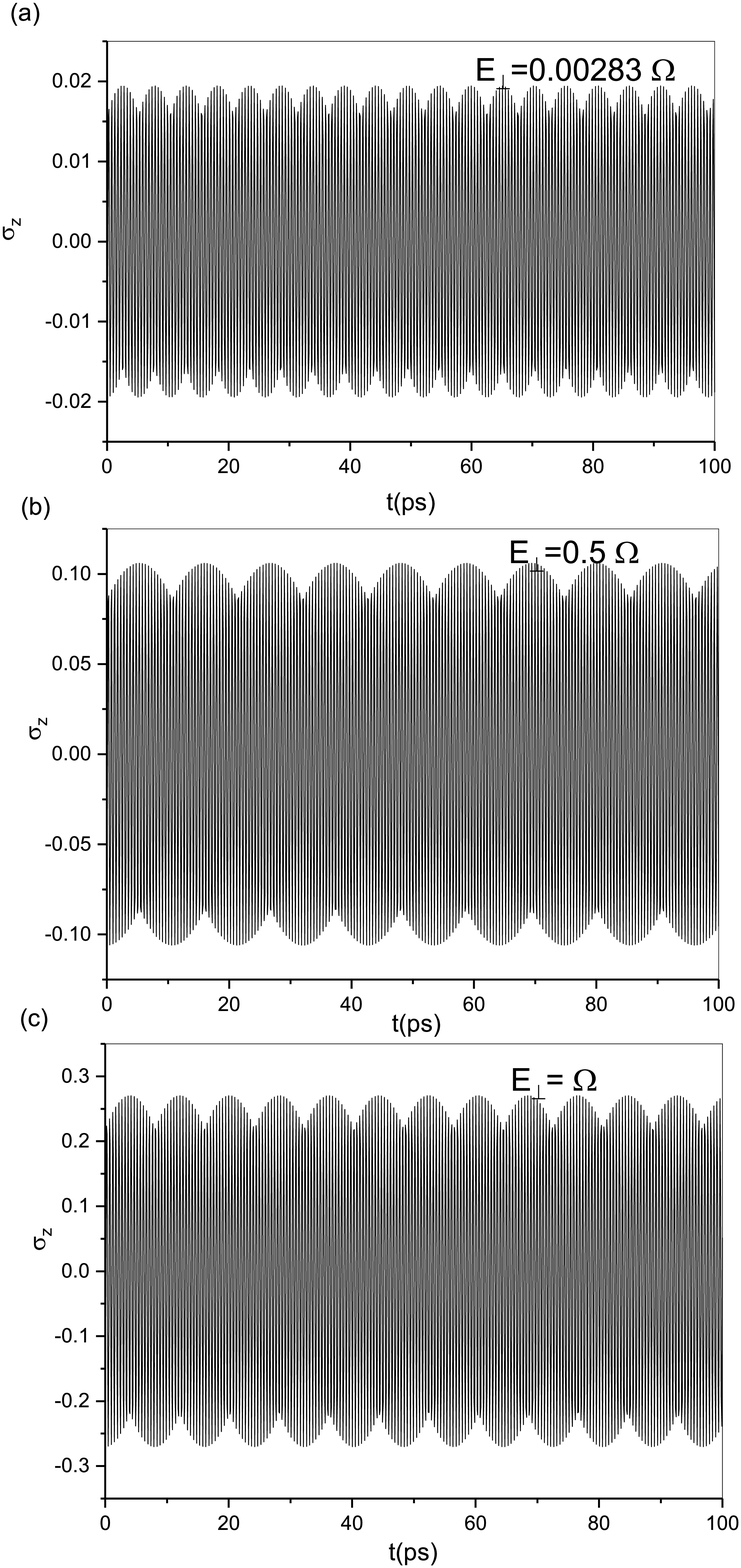}
\label{fig:side:b}
\end{minipage}
}\\
\subfigure{
\begin{minipage}[t]{0.4\textwidth}
\includegraphics[width=1\linewidth]{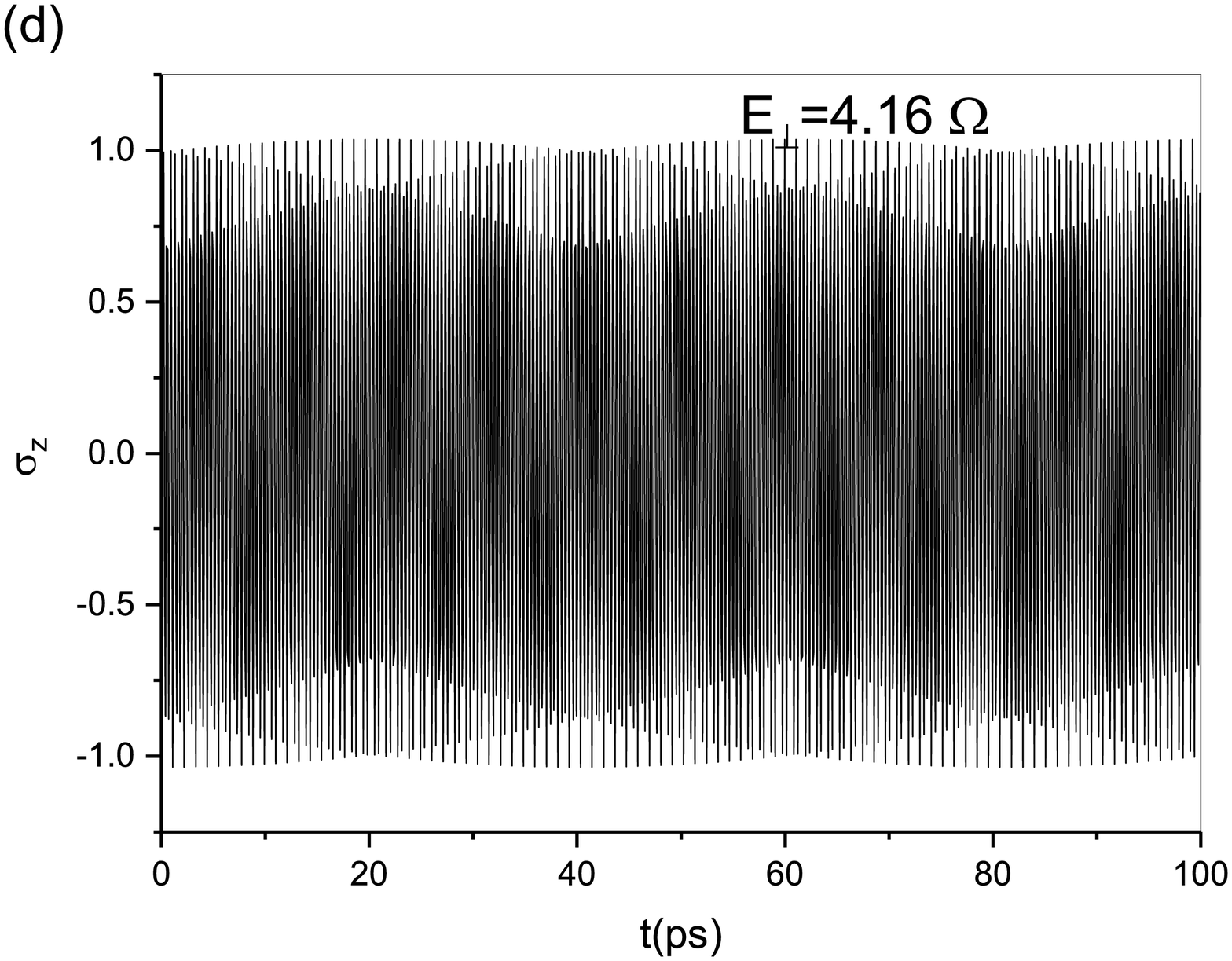}
\label{fig:side:a}
\end{minipage}
}
\caption{Out-of-plane spin polarization of monolayer silicene as a function of time at momentum ${\bf k}=2$ under diffferent strengths of electric field.
The frequency of light used here is $3000$ THz$=7.5$ t.
The corresponding electric field are labeled in the plots.
}
\end{figure}
\clearpage
Fig.6
\begin{figure}[!ht]
   \centering
   \begin{center}
     \includegraphics*[width=0.8\linewidth]{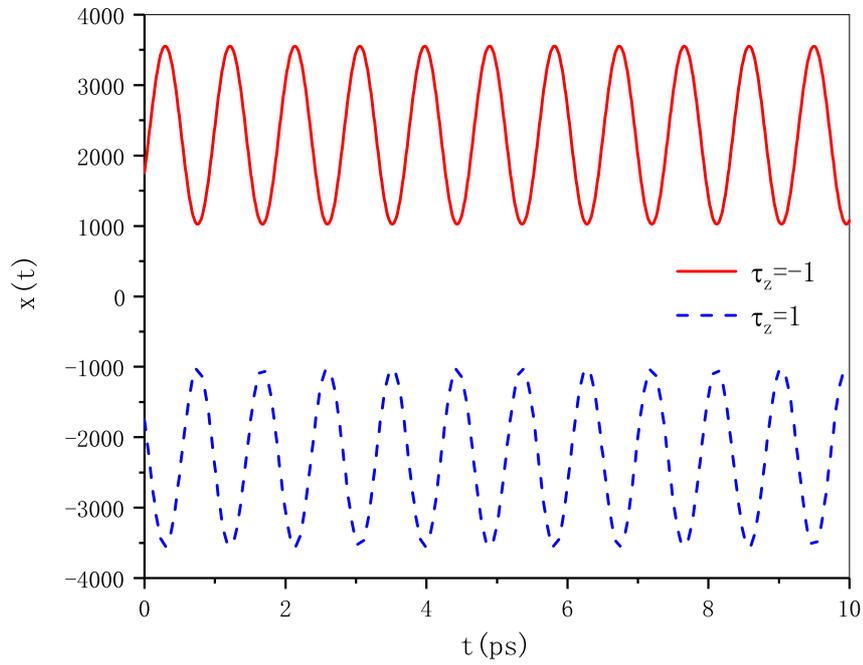}
\caption{Motion of the center of mass of the two pseudospin components ($\tau_{z}=-1$ for upper curve and $\tau_{z}=1$ for lower curve)
obtained by the dipole method.
The frequency of light used here is $3000$ THz$=7.5$ t and 
the corresponding electric field is $E_{\perp}=\Omega$.
The horizontal axis is time in unit of ps, and the vertical axis is in arbitrary unit.
}
   \end{center}
\end{figure}

\end{document}